\documentclass{appolb}
\usepackage{graphicx,color,mathrsfs,amsmath,amssymb}
\usepackage[section]{placeins}
\newcommand{\reels}{\mathbb{R}}
\begin{document}
\title{Towards finite field theory: the Taylor-Lagrange regularization scheme
\thanks{Presented at Light Cone 2012, Krakow, Poland, 8-13 July 2012.}
}
\author{Jean-Fran\c cois Mathiot
\address{Clermont Universit\'e, Laboratoire de Physique Corpusculaire, BP10448,  63000 Clermont-Ferrand, France}
}
\maketitle
\begin{abstract}
We recall a natural framework to deal with local field theory in which bare amplitudes are completely finite. We first present the main general properties of this scheme, the so-called Taylor-Lagrange regularization scheme. We then investigate the consequences of this scheme on the calculation of perturbative radiative corrections to the Higgs mass within the Standard Model. Important consequences for the renormalization group equations are finally discussed.
\end{abstract}
  
\section{Introduction}
The experimental tests of the standard model of particle physics are entering a completely new era with the first $pp$ collisions at LHC (CERN) in the TeV energy range. In a bottom-up type approach, any experimentally verified deviation above some energy scale $\Lambda_{eff}$ from the theoretical predictions within the Standard Model will be a sign of new physics.  
In any physical process, the requirement of theoretical consistency  demands that any characteristic intrinsic momentum which is relevant for the description of any physical  process  should be less then $\Lambda_{eff}$.  If this is not the case, the Standard Model Lagrangian, ${\cal L}_{SM}$,  should be supplemented by effective operators of dimension $(mass)^{i+4}$, with $i>0$, compatible with the symmetries of the system. For a given physical process, these new contributions are proportional to $\left(\Lambda_k/\Lambda_{eff}\right)^i$, where $\Lambda_k$ is any of these characteristic intrinsic momentum. 

At tree level, the momentum $\Lambda_k$ is defined by the typical kinematical variables of the process. It is thus completely under  control. However, beyond tree level, one has to deal with internal momenta in loop contributions that may be large. In that case, this physical intrinsic scale should not be mixed up with spurious scales originating from the possible divergence of bare amplitudes.

In this study, we shall focus on the Taylor-Lagrange regularization scheme (TLRS) developped in Ref.~\cite{GW}. This scheme originates from the well known observation that the divergences of bare amplitudes can be traced back to the violation of causality, originating from  ill-defined products of distributions  at the same point \cite{Scharf,aste}. The correct mathematical treatment, known since a long time,  is to consider covariant fields
as operator valued distributions (OPVD), these distributions being applied on test functions with well-defined mathematical properties. These considerations lead to the TLRS \cite{GW,grange, GW_gauge}. 
 Since this scheme is completely finite, by construction, it is not plagued with unphysical large scales originating from divergent  integrals. 
 
\section{Construction of the physical fields}
Any  quantum field $\phi(x)$ - taken here as a scalar field for simplicity -
 should be considered as an OPVD. It is given  by a distribution, $\phi$, which defines a functional, $\Phi$, with respect to a test function 
$\rho$ according to 
$\Phi(\rho) \equiv \int d^4y \phi(y) \rho(y)$.
The physical field $\varphi(x)$ is then defined in terms of the translation, $T_x$,  of $\Phi(\rho)$, given 
by
\begin{equation} \label{conv}
\varphi(x) \equiv T_x \Phi(\rho) =\int d^4y \phi(y) \rho(x-y).
\end{equation}
The test function $\rho$ should belong to the Schwartz space $\mathscr{S}$ of fast decrease functions at infinity. This property insures that the physical field $\varphi(x)$ is a continuous function - as well as all its derivatives - and is solution of the Klein-Gordon equation. 

We shall consider a test function $\rho$ with a typical spatial extension $a$ (in each space-time dimension). If we demand that the effective Lagrangian we start from remains local, we should consider the limit $a \to 0$. This is analoguous to the continuum limit in lattice gauge calculations. In practice, it is enough to demand that $a$ is sufficiently small, noted by $a \sim 0$, so that physical observables are independent of the particular choice of $\rho$. 
The test function can thus be characterized by $\rho_a(x)$ and the physical field in (\ref{conv}) by $\varphi_a(x)$. In the limit $a \to 0$, we shall  have a priori $\rho_a(x) \to \rho_\eta(x)$ and hence $\varphi_a(x) \to\varphi_\eta(x)$, where $\eta$ is an arbitrary, dimensionless, scale since in the limit $a \to 0$, we also have $a/\eta \to 0$, with $\eta >1$. 

For practical calculations, it is convenient to construct physical fields in momentum space. If we denote by $f_\eta$ the Fourier transform of the test function $\rho_\eta(x)$, we can write ${\varphi_\eta}(x)$ in terms of creation and destruction operators, leading to \cite{GW}
\begin{equation} \label{fk}
\!\varphi_\eta (x)\!=\!\!\int\!\frac{d^3{\bf p}}{(2\pi)^3}\frac{f_\eta(\varepsilon_p^2,{\bf p}^2)}{2\varepsilon_p}
\left[a^\dagger_{\bf p} e^{i{p.x}}+a_{\bf p}e^{-i{p.x}}\right],
\end{equation}
with  $\varepsilon^2_p = {\bf p}^2+m^2$.  
It is apparent from this decomposition that test functions should be attached to each fermion and boson fields. Each 
propagator being the contraction of two fields should be proportional to $f_\eta^2$. In order to have a dimensionless argument for $f_\eta$, which is also dimensionless, 
we shall introduce an arbitrary scale $\Lambda$ to "measure" all momenta. 

Note that the condition $a \sim 0$ implies, in momentum space, that $f_\eta$  is constant almost everywhere, which we shall denote by $f_\eta \sim cte$. It is sufficient to consider such a constant equal to $1$ in order to conserve the normalization of the field and to have the property $T_x \varphi_\eta (x)=\varphi_\eta (x)$. 

The function $f_\eta$ belongs also to the space $\mathscr{S}$, with infinite support. To construct it from a practical point of view, we shall start from a sequence  of functions, denoted by $f_\alpha$, with compact support, and build up from a partition of unity (PU) \cite{grange}. This function is thus zero outside a finite domain of $\reels^4$, along with all its derivatives (super-regular function). 
The parameter $\alpha$, chosen for convenience between 0 and 1, controls the lower and upper limits of the support of $f_\alpha$.  

\section{Construction of (finite) extended bare amplitudes}
Any  amplitude associated to a singular distribution $T(X)$, written schematically as
\begin{equation} \label{cala}
{\cal A}_\alpha = \int_0^\infty dX \ T(X) \ f_{\alpha}(X),
\end{equation}
for a one dimensional variable $X$ for simplicity, is, from the properties of a PU, independent of the precise choice of $f_\alpha$ \cite{GW}.  We shall detail here for shortness only ultra-violet extensions. 

We must now verify that in a given limit the function $f_\alpha$ is equivalent to the fast decrease function $f_\eta$.
For that, we shall verify that the amplitude $A_\eta = \lim_{\alpha \to 1^-}A_\alpha$ is independent of the upper boundary of the support of the test function $f_\alpha$, denoted by  $X_{max}$. It is easy to see that with a naive construction of $f_\alpha$, using a sharp cut-off at $X_{max}$ for instance, this constraint is not verified.

Following Ref.~\cite{GW}, we shall consider a running boundary $H_\alpha(X)$ defined in the UV domain by
\begin{equation} \label{running}
f_\alpha(X \ge H_\alpha(X)) = 0 \ \ \ \mbox{ for} \ \ \ \ H_\alpha(X)\equiv \eta^2 X g_\alpha(X) + cte,
\end{equation}
where $\eta$ is an arbitrary dimensionless scale which should only be larger than $1$.
The function $g_\alpha(X)$ is chosen so that when $\alpha \to 1^-$, $X_{max}$ defined by $X_{max}=H_\alpha(X_{max})$ goes to infinity. A typical example of $g_\alpha(X)$ is given by 
$g_\alpha(X)=X^{\alpha-1}$.
In the limit $ \alpha \to 1^-$, we have $g_\alpha (X) \to 1^-$ except in the asymptotic region  $X \sim X_{max}$.
Note that this running boundary also guaranties the scale invariance already mentioned in the construction of the test function in coordinate space. 
This condition is equivalent to having an ultra-soft cut-off \cite{GW}, i.e. an infinitesimal drop-off of the test function in the asymptotic region, the rate of drop-off being governed by the arbitrary scale $\eta$. 

With this condition, the TLRS proceeds as follows.
Since $f_\alpha$ is a super-regular function, it is  equal to its Taylor remainder to any order $k$. We can thus apply the following  Lagrange formula to $f_\alpha$, after separating out for convenience an intrinsic scale $\lambda$ from the (running) dynamical variable $X$. 
\begin{equation} \label{faX}
f_\alpha(\lambda X)=-\frac{X}{\lambda ^k k!}\int_\lambda ^\infty\frac{dt}{t} (\lambda -t)^k \partial_X^{k+1}\!\left[ X^k f_\alpha(Xt)\right].
\end{equation}
This Lagrange formula is valid for any order $k$, with $k\ge0$.
Starting from the general amplitude $A_\alpha$ written in (\ref{cala}),  and after integration by part, with the use of (\ref{faX}), we get
\begin{equation} \label{afin}
{\cal A}_\alpha = \int_0^\infty dX \ \widetilde T^>_\eta (X) f_\alpha(X).
\end{equation}
In the limit $f_\alpha \to 1$, i.e. for $\alpha \to 1^-$, we have \cite{GW} 
\begin{equation} \label{Tex}
\widetilde T_\eta^>(X)\equiv\frac{(-X)^{k}}{\lambda^k k!} \partial_X^{k+1} \left[ X T(X)\right] \int_\lambda^{\eta^2} \frac{dt}{t} (\lambda-t)^k.
\end{equation}
This is the so-called extension of the singular distribution $T(X)$ in the UV domain. The value of $k$ in (\ref{Tex}) corresponds to the order of singularity of the original distribution $T(X)$ \cite{GW}. In the limit $\alpha \to 1^-$,  the integral over $t$ is independent of $X$ with the choice (\ref{running}) of a running boundary,  while the extension of $T(X)$ is  no longer singular due to the derivatives in (\ref{Tex}). We can therefore safely perform the limit $\alpha \to 1^-$ in (\ref{afin}), and get
\begin{equation}
{\cal A}_\eta = \int_0^\infty dX \ \widetilde T_\eta^>(X) ,
\end{equation}
which is well defined but depends on the arbitrary dimensionless scale $\eta$. This scale is the only remnant of the presence of the test function. 
For massive theories with a mass scale $M$, it is easy to translate this arbitrary dimensionless scale $\eta$ to an arbitrary "unit of mass" $\mu=\eta M$. For massless theories, one can identify similarly an arbitrary unit of mass $\mu = \eta \Lambda$. This unit of mass is analogous to the well known, and also arbitrary, unit of mass of dimensional regularization (DR). 
Note that we do not need to know the explicit form of the test function in the derivation of the extended distribution $\widetilde T^>_\eta(X)$. 
We only rely on its mathematical properties  and on the running construction of the boundary conditions. 

\section{Application of radiative corrections in the Higgs sector}
\subsection{The fine-tuning problem revisited}
Using a na\" ive cut-off  to regularize the bare amplitudes, the (square of the) physical mass of the Higgs particle, denoted by $M_H$  can be schematically written as
\begin{equation} \label{fine}
M_H^2 = M_0^2 + b \ \Lambda_C^2 + \ldots,
\end{equation}
where $M_0$ is the mass parameter of the Higgs particle in the bare effective Lagrangian, and $b$ is a combination of the top quark, $W,Z$ bosons and  Higgs masses. The so-called fine-tuning problem arises if one wants to give some kind of physical reality to the bare mass $M_0$. Since $\Lambda_C$ should be much larger than any characteristic energy scale relevant for the description of  the theoretical physical amplitude,  a large cancellation between $M_0^2$ and $b\,  \Lambda_C^2 $ should be enforced by hand --- hence the name fine-tuning --- unless $b$ is zero (the so-called Veltman condition). 

Apart from the question of identifying the magnitude of $\Lambda_C$,  one may come back to the very origin of the fine-tuning problem, i.e. to the divergences of Feynman amplitudes in the standard approach. Within a finite regularization scheme like TLRS, the interpretation of radiative corrections to the Higgs mass is of a very different nature. As we shall see below, the only relevant momentum scales in TLRS are of the order of the Higgs mass, or of the kinematical experimental conditions. There is therefore no fine-tuning problem to worry about.
 
In leading order of perturbation theory, the radiative corrections to the Higgs mass in the Standard Model gives rise to self-energy type corrections according to
\begin{equation} \label{moh}
M_H^2=M_0^2+\Sigma(M_H^2).
\end{equation}
The calculation of the various contributions to the self-energy is very easy in TLRS. Let us illustrate the calculation of the simple Higgs loop contribution. In Euclidean space one has
\begin{equation} \label{f1b}
-i\Sigma_{1b,H}=-\frac{3iM_H^2}{2v^2}\int_0^\infty \frac{d^4k_E}{(2\pi)^4} \frac{1}{k^2_E+M_H^2} f_\alpha\left( \frac{k^2_E}{\Lambda^2} \right),
\end{equation}
where $k^2_E$ is the square of the four-momentum $k$. The test function $f_\alpha$  provides the necessary (ultra-soft) cut-off in the calculation of the integral. Following the lines recalled above, the extended bare amplitude is completely finite and depends on the arbitrary scale $\eta$. It reads \cite{Higgs}
\begin{equation} \label{TLRS1}
\Sigma_{1b,H}=-\frac{3M_H^4}{32 \pi^2v^2} 
\int_0^\infty dX \partial_X \left( \frac{X}{X+1}\right) \int_1^{\eta^2} \frac{dt}{t} 
=- \frac{3M_H^4}{32 \pi^2 v^2} \mbox{ln}\left(\eta^2 \right).
\end{equation}
For completeness, we recall below the result of the direct calculation of (\ref{f1b}) in DR
\begin{equation} \label{DR1}
\Sigma_{1b,H}^{DR}=\frac{3M_H^4}{32 \pi^2v^2} 
\left[ -\frac{2}{\varepsilon} + c  - \mbox{ln} \left( \frac{\mu^2}{M_H^2} \right) \right],
\end{equation}
where $c=\gamma_E-1-\mbox{ln} 4\pi$ and $\gamma_E$ is the Euler constant.
We can already see from these results that  TLRS and DR lead to a similar  scale-dependent logarithmic term, with the identification $\eta^2=\mu^2/M_H^2$. They both depend on a completely arbitrary constant. 

\subsection{Physical scales}
\begin{figure}[btph]
\begin{center}
\includegraphics[width=20pc]{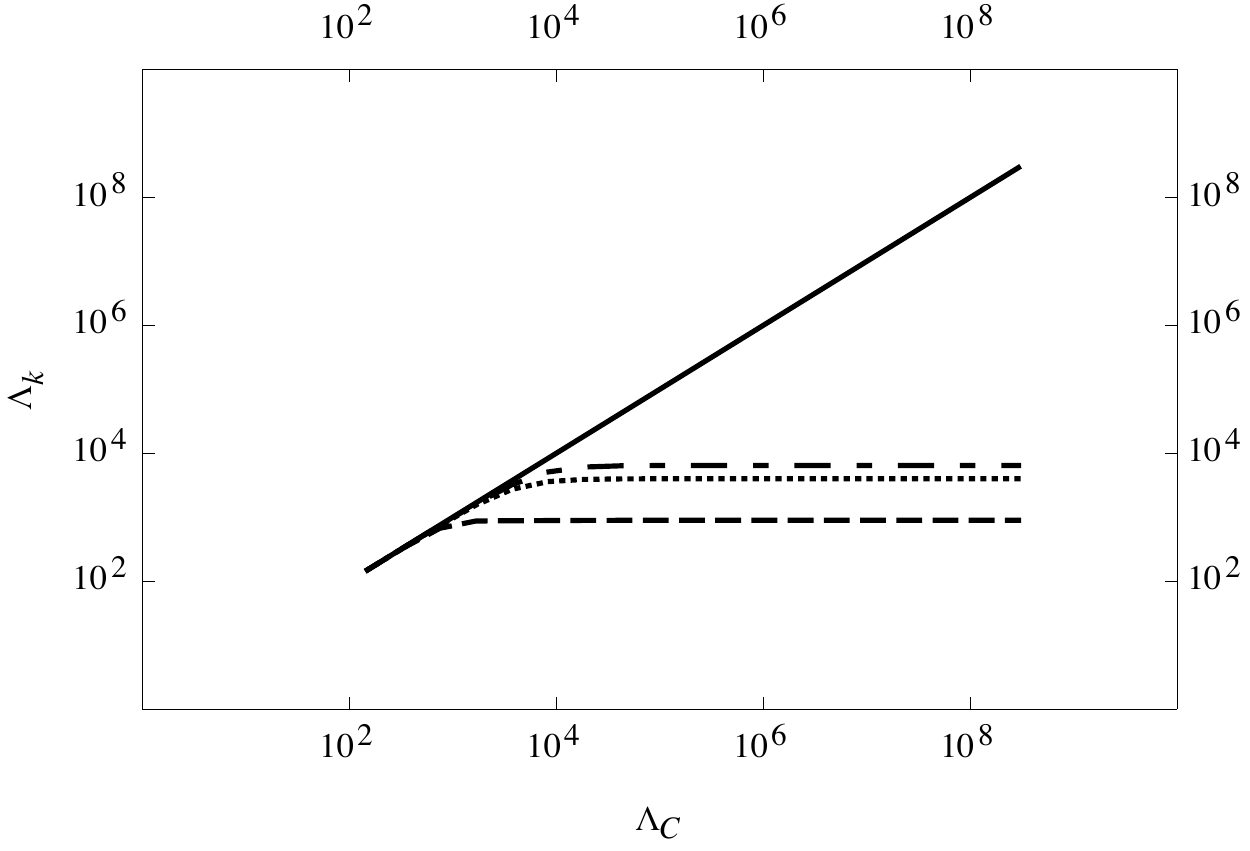}
\caption{Characteristic momentum scale $\Lambda_k$ calculated from the self-energy contribution $\bar \Sigma(M_H^2)$, in two different regularization schemes:   with a na\" ive cut-off (solid line) and using TLRS (dashed line). The
calculation is done for $M_H=125$ GeV, with  $\eta^2=100$. We also show on this figure $\Lambda_k$ calculated
with the  fully renormalized self-energy (\ref{fullR}) for $p^2=-10 \ M_H^2$ (dotted line) and $p^2=-100 \ M_H^2$ (dash-dotted line).  \label{finescale}}
\end{center}
\end{figure}

We shall concentrate in this subsection on the characteristic intrinsic momentum scale $\Lambda_k$ relevant for the calculation of the radiative corrections to the mass of the Higgs particle.
In order to determine $\Lambda_k$ from a quantitative point of view, we shall proceed in the following way. Writing  the self-energy as
$\Sigma(p^2)=\int_0^{\Lambda_C^2}dk_E^2\  \sigma(k_E^2, p^2)$,
we shall define the characteristic momentum $\Lambda_k$ by requiring that the reduced self-energy defined by
$\bar \Sigma(p^2)=\int_0^{\Lambda_k^2}dk_E^2 \ \sigma(k_E^2, p^2)$
differs from $\Sigma(p^2)$ by $\epsilon$ in relative value, i.e. with the constraint 
${\bar \Sigma(p^2)}/{\Sigma(p^2)}=1-\epsilon$,
provided we have $\vert \bar \Sigma(p^2) \vert < \vert \Sigma(p^2) \vert$.
In the Standard Model, $\epsilon$ can be taken of the order of $1 \%$. 
We show in Fig.~\ref{finescale} the characteristic scale $\Lambda_k$ calculated for two typical  expressions of the self-energy of the Higgs particle, as a function of $\Lambda_C$. The first expression is the bare one given by $\Sigma(M_H^2)$ in (\ref{moh}), while the second one is the  fully (on-shell) renormalized amplitude, i.e. with both mass and wave function renormalization, defined by 
\begin{equation} \label{fullR}
\Sigma_R(p^2)= \Sigma(p^2)- \Sigma(M_H^2)-(p^2-M_H^2) \left. \frac{d\Sigma(p^2)}{dp^2}\right |_{p^2=M_H^2}
\end{equation}
and calculated at two different values of $p^2$, $p^2=-10 \ M_H^2$ and $p^2=-100 \ M_H^2$. 

The results indicated in Fig.~\ref{finescale} exhibit two very different behaviors. If one considers first the calculation of the bare amplitude, the use of a na\" ive cut-off regularization scheme does not allow to identify any characteristic momentum $\Lambda_k$. Since  $\Lambda_k$ is always very close to $\Lambda_C$, all momentum scales are  involved in the calculation of the bare self-energy. This is indeed a trivial consequence of the fact that the  bare amplitude is divergent  in that case.  However, using TLRS, we can clearly identify a characteristic momentum $\Lambda_k$, since it reaches a constant value for  $\Lambda_C$ large enough. Note also that in this regularization scheme, we can choose a value of $\Lambda_C$ which is arbitrary, as soon as it is much larger than any mass or external momentum of the constituents.  It can even be infinite, since it does not have any physical meaning.
It is in full agreement with the local character of the effective Lagrangian ${\cal L}_{eff}$, since in that case $\Lambda_C$ should be taken to be infinite.

If we consider now the characteristic momentum scale relevant for the description of the fully renormalized amplitude $\Sigma_R$, we can also identify a finite value for $\Lambda_k$ since it saturates at sufficiently large values of $\Lambda_C$ compared to the typical masses and external momenta of the system. This behavior is extremely  similar to the result obtained in the above analysis of the bare amplitude $\Sigma$ using TLRS. This is again not surprising since the fully renormalized amplitude is also completely finite.  It depends only slightly on the external kinematical condition $\Lambda_Q$ (given here by $\sqrt{-p^2}$). In any case, the characteristic momentum scale  is of the order of $\Lambda_Q$, and, what is more important, it is independent  of $\Lambda_C$. One can check  that $\Sigma_R$ is  of course identical in all renormalization schemes.

\section{Final remarks}
\subsection{Interest in light-front dynamics}
The use of the TLRS in light-front dynamics is very natural. Starting from a Fock space expansion of the state vector according to 
$\Phi(p) = \sum_n \Gamma_n(k_1 \ldots k_n) \vert n \rangle$,
with obvious notations, the properties of the test functions are now embedded in the vertex functions $\Gamma_n$ with the replacement
\begin{equation}
\Gamma_n(k_1 \ldots k_n) \to \bar \Gamma_n(k_1 \ldots k_n)=\Gamma_n(k_1 \ldots k_n)f({\bf k_1}^2/\Lambda^2)\ldots f({\bf k_n}^2/\Lambda^2)
\end{equation}
It is a completely nonperturbative implementation of the TLRS. All amplitudes calculated in light-front dynamics will thus be finite, and depend on the arbitrary scale $\eta$, as shown in Ref.~\cite{grange}.

\subsection{Renormalization group equations}
Since all amplitudes do depend a priori on the arbitrary scale $\eta$ embedded in the test function $f_\eta$, all field strengths, bare masses and bare coupling constants do depend on this arbitrary scale also. However, all physical masses and coupling constants, and more generally all physical observables should not depend on $\eta$. We can thus derive a renormalization group equation related to  this invariance.

Since the relation between the ($\eta$-dependent) bare parameters and the ($\eta$-independent) physical ones is mass-dependent, the renormalization group equations will also be mass-dependent, in contrast to DR regularization in the MS scheme. In this latter case, the mass-independence of the renormalization group equations originates from the assumption that bare parameters are independent of the unit of mass inherent to DR. This is at variance with TLRS where the bare parameters do depend on $\eta$. In view of the close relationship we found between $\eta$ and the unit of mass $\mu$ in DR, one may question this assumption. In particular, since the Lagrangian is rather a {\it density} Lagrangian, it may depend a priori on the dimension of space-time, i.e. on $\mu$ also.

\subsection{How and why to use Taylor-Lagrange regularization scheme}
Since physical observables should be independent on the regularization/renormalization schemes which we use to perform explicit calculations, one may wonder how and why to use the Taylor-Lagrange regularization scheme. The first and most evident advantage is that we stay all the time in our physical world! From two different points of view: the dimension of our space-time is the physical four-dimensional space, while all momenta which are not forbidden by kinematical constraints are retained (Nature knows nothing about cut-off's!). Moreover, we do not need to rely on auxiliary fields like Pauli-Villars fields with very large masses.
In standard calculations in perturbation theory, this avoids all complications necessary to treat chiral transitions, related to the definition of $\gamma_5$, or to inforce supersymmetry in arbitrary space-time dimensions. 

 In nonperturbative calculations, the use of TLRS is very natural, like for instance in light-front dynamics. One may expect that this scheme may also shed some light in lattice gauge calculations \cite{lattice}. It does not rely also on any infinite mass limit which becomes very difficult to handle numerically.
 
 While explicit calculations at a fixed order in perturbation theory should be identical in all schemes, the use of renormalization-group improved calculations, where partial resummation of a class of Feynmann diagrams is performed, may lead to quite different results due to the different nature (mass dependence or independence) of the renormalization group equations.

\end{document}